\documentclass[10pt,journal,compsoc]{IEEEtran}
\usepackage[T1]{fontenc}
\usepackage{amsmath,amssymb,amsthm,bm}
\usepackage{graphicx}
\usepackage{booktabs,array}
\usepackage[nocompress]{cite}
\usepackage{url}
\newtheorem{theorem}{Theorem}
\newtheorem{proposition}{Proposition}
\theoremstyle{definition}
\newtheorem{definition}{Definition}
\newtheorem{assumption}{Assumption}
\newcommand{\wv}{\mathbf{w}}
\newcommand{\phiv}{\bm{\phi}}
\newcommand{\Tin}{T_{\mathrm{in}}}
\newcommand{\Tgen}{T_{\mathrm{gen}}}
\newcommand{\Tctx}{T_{\mathrm{ctx}}}

\begin{document}

\title{Token-as-a-Utility: A Model-Agnostic Architecture for Metering and Provisioning AI Inference Services}

\author{Yichiet~Aun,
Soung-Yue~Liew,
Yu~Liang~Lai,
Jing~Jie~Tan,
Dong~Chen,
and~Purnomo~Husnul~Khotimah%
\IEEEcompsocitemizethanks{\IEEEcompsocthanksitem Y.~Aun, S.-Y.~Liew and Y.~L.~Lai are with the Faculty of Information and Communication Technology, Universiti Tunku Abdul Rahman, Kampar, Malaysia.
\IEEEcompsocthanksitem J.~J.~Tan is with the Saw Swee Hock School of Public Health, National University of Singapore, Singapore.
\IEEEcompsocthanksitem D.~Chen is with the School of Physics and Electronics, Nanning Normal University, Nanning, China.
\IEEEcompsocthanksitem P.~H.~Khotimah is with the Research Center for Data and Information Sciences, National Research and Innovation Agency (BRIN), Indonesia.
\IEEEcompsocthanksitem This work has been submitted to the IEEE for possible publication. Copyright may be transferred without notice, after which this version may no longer be accessible.
\IEEEcompsocthanksitem E-mail: \{aunyc, syliew, yllai\}@utar.edu.my, tanjingjie@nus.edu.sg, chendong@nnnu.edu.cn, purn005@brin.go.id. Corresponding author: Yichiet Aun.}}

\markboth{IEEE Transactions on Services Computing}%
{Aun \MakeLowercase{\textit{et al.}}: Token-as-a-Utility}

\IEEEtitleabstractindextext{%
\begin{abstract}Token-based subscriptions are emerging as a commercial model for large language model inference, but extending them into interoperable utility services requires comparable consumption accounting. Raw token counts do not consistently measure inference resource use because computational demand varies with model architecture, context length, tokenisation, and concurrent serving conditions. This paper presents Token-as-a-Utility, a model-agnostic architecture integrating inference metering, service-level-aware provisioning, and inter-operator settlement. We prove that, under batched serving, no per-request meter can be additive, independent of co-resident traffic, and exactly cost-reflective at once. This motivates separating provisioned capacity, metered consumption, and contracted quality into distinct service products. Consumption is expressed in Reference Inference Units (RIUs), derived from prefill and decode resource requirements, while occupancy-indexed operator ratings recover shared serving costs without altering the common unit. Versioned inference records support auditing and re-rating across operators, while software-enforced capacity leases and quality-constrained routing enable provisioning on shared accelerators. The resource model is validated through measurements on three NVIDIA RTX 5000 Ada GPUs, followed by multi-tenant evaluation using a trace-driven model calibrated to them. For the reference model, occupancy-indexed rating limits statement energy error to 9\% under measured traffic, compared with up to 70.5\% under fixed-occupancy rating. In calibrated multi-tenant experiments, tenant mispricing decreases from 7.36 times under flat token counting to 2.39 times with RIUs and 1.06 times after occupancy-based reconciliation. These findings demonstrate how portable inference accounting can be combined with load-dependent cost recovery and contract-based provisioning to support heterogeneous inference services across models and operators.
\end{abstract}

}

\maketitle
\IEEEdisplaynontitleabstractindextext
\IEEEpeerreviewmaketitle

\section{Introduction}\label{sec:intro}

\IEEEPARstart{L}{arge} language model (LLM) inference has acquired the commercial form of a utility. China's National Data Administration reported that daily token calls exceeded 140 trillion by March 2026, against about 100 billion in early 2024 \cite{xinhua_nda_tokens_2026}. The three Chinese mobile operators sell monthly token allowances that resemble mobile data bundles, and China Telecom adds connectivity and security to the same subscription \cite{gsmai2026token}. Singtel launched a token-metered managed inference service in October 2026 \cite{singtel_reai_taas_2026}. A likely next product bundles a broadband rate, say 100~Mbit/s, with an inference entitlement that applies whichever model answers.

A broadband rate can be sold without naming the router, because a bit is a fixed quantity of information and a bit rate is a reservable capacity. Raw token counts have neither property. The same sentence yields different counts under different tokeniser vocabularies, and the work behind a token depends on architecture, context length and the number of requests sharing each decode step. Raw token counts therefore do not provide a model-independent measure of inference work.

This paper builds on three lines of prior work: cross-model units such as Google's burndown rates and Azure's normalized tokens \cite{google2026gsu,azure2026ptusizing,alibaba2026tokenplan}, batching-sensitive attribution of serving energy \cite{luo2026jouleshare}, and token-based fairness such as the Virtual Token Counter (VTC) \cite{sheng2024fairness,khan2024fairserve}. The vendor units apply within one seller's boundary, without a published derivation, error bound or procedure for two sellers to settle in the same unit. Network--compute standards call for normalised compute metrics but leave the method open \cite{rfc10053,itu2026y2504}, and cross-model signals serve autoscaling rather than billing \cite{zhang2026crossmodel}.

This paper presents Token-as-a-Utility, an operational framework for metering, contracting and settling heterogeneous inference services when resource cost depends on batching. It separates a portable reference quantity, the RIU, which never changes with batch occupancy (the number of requests decoded together), from occupancy-dependent operator ratings, a separate adjustment for the conditions under which an operator delivers it. The framework is model-agnostic at the service and accounting interface, where every model is billed in the same unit, while model-specific coefficients, published per model and vintage, convert each request's features into that unit. Per-request records keep the features behind both quantities, so a bill can be audited, re-rated and settled between operators. The contributions are as follows.
\begin{enumerate}
\item \emph{A service architecture} (Section~\ref{sec:arch}) that sells capacity, consumption and quality as separate products, motivated by an impossibility result for per-request metering under batch sharing.
\item \emph{A metering method} (Section~\ref{sec:meter}) that derives Reference Inference Units (RIUs) from the prefill and decode resource model, with occupancy-indexed operator ratings, declared error bounds (accuracy classes), a bound on gains from repackaging and a re-rating rule between coefficient versions (vintages) that is exact when records carry the feature vector (Proposition~\ref{prop:rerate}).
\item \emph{Provisioning without hardware partitioning} (Section~\ref{sec:prov}) through a software capacity lease, a declared contention ratio and routing that checks legality first.
\item \emph{An evaluation calibrated and validated on hardware} (Section~\ref{sec:eval}) at Universiti Tunku Abdul Rahman (UTAR). Serving efficiency, energy, mixture-of-experts weight traffic, swap times and latency under drawn leases are measured against decision rules fixed in advance; the multi-tenant studies, including settlement between operators (Section~\ref{sec:fed}), run on a trace-driven model calibrated with these measurements. Each result is labelled measured, derived, calibrated model or proposed.
\end{enumerate}

\section{Background, Requirements and Related Work}\label{sec:related}

\emph{Utility service models.} Utility computing treats computation, its pricing and its service-level agreements (SLAs) as services problems \cite{rappa2004utility,buyya2009cloud,paleologo2004price,keller2003wsla}. Network pricing separates access capacity, usage and service class \cite{courcoubetis2003pricing}, and utility tariffs recover shared fixed cost through a capacity charge while the per-unit charge follows marginal cost \cite{brown1986theory}. Inference vendors sell reserved throughput \cite{azure2026ptu,google2026gsu}, on-demand tokens and priority tiers \cite{openai2026priority}, each in its own unit.

\emph{Normalised units and attribution.} A standard inference token has been proposed for futures contracts without an operational measurement \cite{xing2026futures}. A commercial energy unit attributes kilowatt-hours to output tokens through a proprietary engine \cite{antarctica2025otm}, and MLPerf Endpoints plans cost and power normalisation for systems serving the same model \cite{mlcommons2026endpoints}. Abstract credits have also spanned resource types and changing fleets \cite{access2026credits,aws_rds_instance_classes_ecu}. Concurrent work estimates cost per token under concurrency \cite{patil2026concurrency}, attributes batch latency with a roofline model \cite{jin2025llmvisor}, measures how request energy depends on batching \cite{vellaisamy2026energy,delavande2026efficiency}, prices batching externalities and latency tiers \cite{yao2026costalignment,mcdougall2026pricingtime}, and compares token-proportional with Shapley attribution \cite{luo2026jouleshare,urista2026kvcache,yuan2026wattsonllm}. They leave open a vendor-neutral consumption table with a declared accuracy class and a settlement procedure.

\emph{Serving and service-level control.} Continuous batching and paged key--value (KV) cache memory let one accelerator serve many tenants \cite{yu2022orca,kwon2023vllm}; prefill (prompt processing) is compute-bound and decode (output generation) bandwidth-bound \cite{zhong2024distserve}; and queueing aware of service-level objectives (SLOs), multi-model pooling, latency isolation and fairness defined over resource cost are well studied \cite{chen2025slosserve,patke2024qlm,xiang2025aegaeon,bali2026fairinference,wei2025equinox,yang2025justitia}. Clients find provider-side token inflation and model substitution hard to detect \cite{hoque2026inflation,cai2025substitution,zhang2026iris}; this work adds the contracts these mechanisms enforce.

\emph{Tokenisation disparity.} Tokenised length of identical text differs by up to fifteen times across languages \cite{petrov2023tokenizers}, at a cost borne by speakers of under-served languages \cite{ahia2023languages,srivastava2026tokenizertax}; this paper connects that disparity to a cross-model billing unit.

\emph{Cost allocation and settlement.} Budget balance, additivity and consistency single out Aumann--Shapley prices for shared facilities \cite{billera1982allocation,mirman1982demand}; on cost functions homogeneous of degree one the same axioms give marginal-cost prices \cite{mirman1983homogeneous,haimanko2002marginal,moulin2002axiomatic}. Shapley settlement between network operators is established \cite{ma2010shapley}, and GSMA's federated operator platform exchanges usage of CPU, memory and storage while leaving settlement out of scope \cite{gsma2026opg02}.

\begin{table}[t]
\caption{Requirements of an inference utility and coverage by prior work: $\bullet$ addressed, $\circ$ partly, -- not addressed. V vendor units \cite{google2026gsu,azure2026ptusizing}; F fair schedulers \cite{sheng2024fairness}; C cost models \cite{patil2026concurrency,vellaisamy2026energy}; N network--compute federation \cite{itu2026y2504,gsma2026opg02}; A cost allocation \cite{billera1978telephone,luo2026jouleshare}; T this work.}\label{tab:req}
\centering\footnotesize\setlength{\tabcolsep}{4.5pt}
\begin{tabular}{@{}p{4.55cm}cccccc@{}}
\toprule
Requirement & V & F & C & N & A & T\\
\midrule
R1 comparable consumption across models, vendors & $\circ$ & -- & $\circ$ & -- & -- & $\bullet$\\
R2 published derivation and declared error & -- & -- & $\circ$ & -- & -- & $\bullet$\\
R3 capacity, consumption, quality as products & $\circ$ & -- & -- & $\circ$ & -- & $\bullet$\\
R4 provisioning without hardware partitioning & -- & $\bullet$ & -- & -- & -- & $\bullet$\\
R5 auditable per-request record, dispute rule & -- & -- & -- & $\circ$ & $\circ$ & $\bullet$\\
R6 settlement across operators and versions & -- & -- & -- & $\circ$ & $\circ$ & $\bullet$\\
\bottomrule
\end{tabular}
\end{table}

Selling inference as a metered service across models and operators creates six requirements (Table~\ref{tab:req}). Tenants need comparable consumption (R1) with a checkable derivation and error (R2). Shared batch cost must be recovered without distorting the unit (R3), often without hardware partitioning (R4). Disputes need per-request evidence (R5), and operators on different coefficient versions must settle (R6). No prior line addresses all six together.

\emph{Objectives.} A cross-model meter can target resource fidelity, language equity (equal charges for equal content in every language) or incentive compatibility (no gain from misreporting tokens). This paper targets resource fidelity, so native-token metering retains the token premium of some languages, which Section~\ref{sec:meter} discloses as an attribute of the quality class. Character-based pricing addresses incentive compatibility \cite{velasco2026overcharging}, a complementary objective.

\section{Token-as-a-Utility System Model and Architecture}\label{sec:arch}

\subsection{Three Products}
A utility customer buys a reserved capacity, a metered consumption and a contracted quality of supply, and Token-as-a-Utility keeps these three products. \emph{Provisioned capacity} is a floor rate of service, enforced by scheduler reservations and billed whether or not it is drawn. \emph{Metered consumption} is measured in RIUs, which add up across requests, models, operators and billing periods. \emph{Contracted quality} has two parts. A quality class states which models may serve a request, through floors on capability, latency and tokeniser fertility (tokens per unit of content). A traffic class states when the request is served. Models may substitute for each other only inside the quality class, so the consumption unit need not describe answer quality. Unlike electricity, where $\mathrm{kWh}=\mathrm{kW}\times\mathrm{h}$, a token-rate entitlement converts into reference work only once the model and serving conditions are specified, so a floor of 100 tokens per second is sold as a rate, not a token allowance.

\subsection{Formal Requirement for Separating the Products}
A \emph{meter} is a map $U:(q,\sigma)\mapsto\mathbb{R}_{\ge0}$ from a request $q$ and its serving context $\sigma=(o,m,c)$ (operator, model and configuration) to RIUs. The \emph{ambient state} $\omega$ comprises the co-resident batch, cache contents and queue, with batch occupancy $n(\omega)$, the number of sequences decoded together. A workload $\mathcal{Q}$ consumes realised resource $R(\mathcal{Q},\omega)$, and the operator admits workloads from a declared envelope $\mathcal{E}$. Throughout, \emph{cost} means this resource; in the evaluation it is allocated GPU time. \emph{Price}, the money charged per unit, is set separately by each operator. A metered unit is expected to satisfy (P1) additivity, $U(\mathcal{Q})=\sum_{q\in\mathcal{Q}}U(q)$; (P2) ambient independence, $U(q;\sigma)$ does not depend on $\omega$; and (P3) exact cost-reflectiveness, $\sum_{q\in\mathcal{Q}}U(q)=\lambda_\sigma^{-1}R(\mathcal{Q},\omega)$ for all admissible $(\mathcal{Q},\omega)$ with a rate $\lambda_\sigma>0$ declared in advance and constant in $\omega$.

\begin{assumption}[Batch sharing]\label{ass:H}
Each decode step reads the active weights, $W_b$ bytes, once regardless of the number $n$ of sequences in flight, while each sequence reads its own KV state of $K(t)$ bytes at context $t$, so that the resource per decoded token of one sequence is
\begin{equation}\label{eq:rho}
\rho(t,n)=W_b/n+K(t).
\end{equation}
\end{assumption}

\begin{theorem}[No exact, additive, ambient-independent meter]\label{thm:imposs}
Fix $\sigma$, let Assumption~\ref{ass:H} hold with $W_b>0$, and let $\mathcal{E}$ contain two workloads of identical requests served at occupancies $n_1\neq n_2$. No meter satisfies P1, P2 and P3 on $\mathcal{E}$.
\end{theorem}
\begin{proof}
Let $\mathcal{A}$ serve $n$ identical requests concurrently and $\mathcal{B}$ serve them one at a time, each generating $\Tgen$ tokens at context $t$. By \eqref{eq:rho}, $R(\mathcal{A})=\Tgen(W_b+nK)$ and $R(\mathcal{B})=n\Tgen(W_b+K)$, so $R(\mathcal{B})/R(\mathcal{A})>1$ for $n>1$. By P2 each $U(q)$ is equal in both workloads, so by P1 $U(\mathcal{A})=U(\mathcal{B})$, and P3 with one $\lambda_\sigma$ forces $R(\mathcal{A})=R(\mathcal{B})$, a contradiction.
\end{proof}

Theorem~\ref{thm:imposs} instantiates a classical result: a budget-balanced price per metered unit that does not change when aggregate demand is scaled exists only if cost is homogeneous of degree one, which for one metered quantity means linear cost \cite{mirman1983homogeneous,haimanko2002marginal,moulin2002axiomatic}. Decode cost $\Tgen(W_b+nK)$ is affine in occupancy with a positive intercept. For mixture-of-experts (MoE) models the weight read grows with $n$, but the measured read per sequence still falls as $n$ grows (Section~\ref{sec:hw}), so the result still applies. The theorem adds the serving mechanism that removes homogeneity, its magnitude (one to two orders of magnitude, Section~\ref{sec:eval}) and its design consequence; token-pricing analyses that assume homogeneous production address a different setting \cite{bergemann2025economics}.

\begin{table}[t]
\caption{Tariff designs permitted by Theorem~\ref{thm:imposs}.}\label{tab:designs}
\centering\footnotesize
\begin{tabular}{@{}p{1.85cm}p{1.0cm}p{4.9cm}@{}}
\toprule
Design & Gives up & Consequence and examples\\
\midrule
Tariff metering & P3 & coefficients fixed in advance, operator bears occupancy risk; vendor units \cite{google2026gsu,azure2026ptusizing}\\
Ex-post allocation & P2 & shares of realised cost; a charge depends on co-tenants \cite{billera1978telephone,luo2026jouleshare}\\
Reservation & P1 & accelerator time of a slice; no per-request quantity to settle \cite{azure2026ptu}\\
Separated (this work) & per product & RIU keeps P1, P2 and P3 within a declared error; capacity charge and occupancy-indexed operator rating recover cost\\
\bottomrule
\end{tabular}
\end{table}

Every tariff gives up one property (Table~\ref{tab:designs}). The separated design remains subject to the theorem and combines existing tariff forms: tariff metering for the quantity, a capacity charge for the shared weight stream (the weight read of each decode step) and an ex-post allocation of the residual by occupancy band. Tenants are quoted, budgeted and settled in RIUs, which keep additivity (P1) and ambient independence (P2) and are cost-reflective (P3) within a declared error. The RIU itself never changes with occupancy: the residual cost of delivery is recovered after the fact (the true-up) through a separate operator rating indexed by the recorded occupancy band, so occupancy dependence is confined to the rating and declared on each record. The charges are disjoint: a floor of $F$ tokens per second costs $F\beta$ per second, tokens inside it are billed without the $\beta$ term, and only on-demand tokens carry $\beta$ and its true-up (worked statement in Supplement S1).

\subsection{Architecture}

\begin{figure*}[t]
\centering
\includegraphics[width=\textwidth]{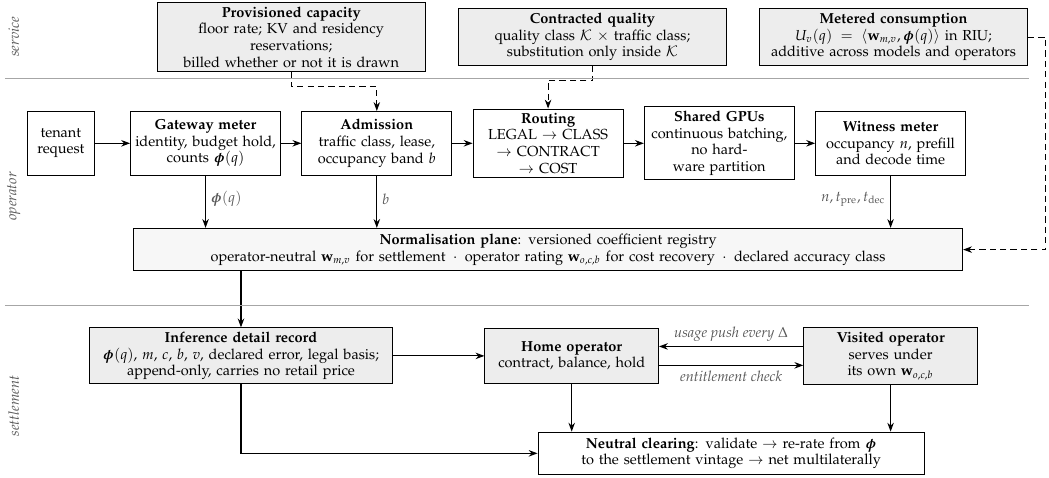}
\caption{Token-as-a-Utility architecture. The three products (top) govern an operator's mechanisms (middle): capacity is enforced at admission, quality at routing, and the normalisation plane computes consumption from gateway counts and engine-recorded serving conditions. The inference detail record (IDR) carries the feature vector $\bm{\phi}(q)$, so settlement can re-rate it to a common coefficient vintage (bottom). Dashed arrows: contractual governance; solid arrows: data flow.}\label{fig:arch}
\end{figure*}

Figure~\ref{fig:arch} shows the architecture. The \emph{gateway meter} authenticates the tenant, takes a budget hold in RIUs and records the token counts that form the feature vector $\phiv(q)$. \emph{Admission} applies the traffic class, the capacity lease and the occupancy band of the accuracy class. \emph{Routing} checks legality, quality class, contract and cost in that order, and dispatches to a replica on \emph{shared accelerators}. A second meter in the serving engine, the \emph{witness meter}, records occupancy, prefill and decode time and preemptions, which the gateway cannot observe; the engine's own metrics carry no tenant label \cite{kwon2023vllm}. The two meters are reconciled per request: token counts must agree exactly, and RIU values within the declared error. The \emph{normalisation plane} holds the versioned coefficient registry. Each request produces one append-only \emph{inference detail record} (IDR) carrying $\phiv(q)$, model, configuration, occupancy band, vintage, declared error and, for cross-border routing, the legal basis. It carries no retail price, because the RIU is a wholesale quantity.

\section{Model-Agnostic Inference Metering}\label{sec:meter}

\subsection{Three Normalisation Problems}

\begin{figure}[t]
\centering
\includegraphics[width=\columnwidth]{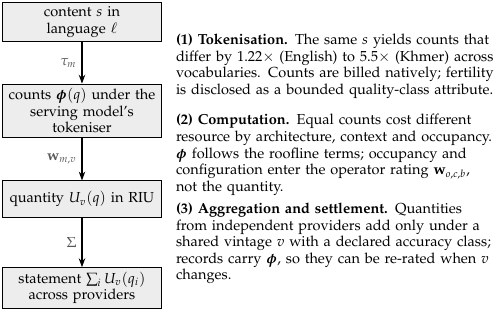}
\caption{The three normalisation problems between content and a settleable quantity, and the mechanism that handles each.}\label{fig:norm}
\end{figure}

Turning heterogeneous inference into one quantity involves three problems (Fig.~\ref{fig:norm}): \emph{tokenisation normalisation}, because the same content yields different counts under different vocabularies; \emph{computational normalisation}, because equal counts consume different resources by architecture, context, precision and occupancy; and \emph{commercial normalisation}, because independent providers can sum, dispute and settle consumption only under a shared, versioned convention with a known error.

\subsection{Resource Model and Reference Inference Unit}
For $\Tin$ fresh input tokens, prefill performs $F_{\mathrm{pre}}=2N_a\Tin+2d_{\mathrm{attn}}[L_g\Tin^2/2+L_\ell(w\Tin-w^2/2)]$ operations, where $N_a$ is the number of active parameters, $L_g$ and $L_\ell$ the numbers of global and sliding-window layers and $w$ the window, for $\Tin\ge w$. Attention counts one multiply--add per query--key pair and dimension, the scaling-law convention (Supplement S1). Over $\Tgen$ decode steps a sequence starting at context $\Tctx$ moves
\begin{equation}\label{eq:decode}
B_{\mathrm{dec}}=\frac{W_b\Tgen}{n}+k_{\mathrm{tok}}\Big(\Tctx\Tgen+\tfrac{\Tgen(\Tgen-1)}{2}\Big)+k_{\mathrm{fix}}\Tgen
\end{equation}
bytes, where $k_{\mathrm{tok}}$ is the KV size per token, computed from key and value heads rather than query heads, and $k_{\mathrm{fix}}$ the per-sequence floor of sliding-window layers. Converting both into time through the device roofline (peak compute and bandwidth) at a declared reference occupancy $n_0$ gives
\begin{equation}\label{eq:riu}
\begin{aligned}
U_v(q)&=\langle\wv_{m,v},\phiv(q)\rangle,\\
\phiv(q)&=\big(1,\Tin,\Tin^2,\Tgen,\Tctx\Tgen+\tfrac{\Tgen^2}{2},N_{\mathrm{img}},V\big),
\end{aligned}
\end{equation}
with coefficients $(c_0,\alpha_1,\alpha_2,\beta,\mu,\varepsilon_{\mathrm{enc}},\alpha_1)$ keyed to model $m$ and vintage $v$, where $N_{\mathrm{img}}$ images contribute $V$ visual tokens that are prefilled, and so rated, like text. The coefficient of $\Tgen$ is $\beta-\mu/2+\kappa_{\mathrm{fix}}$, with $\kappa_{\mathrm{fix}}$ the sliding-window floor in RIUs, so \eqref{eq:riu} reproduces \eqref{eq:decode}; Supplement S1 gives the cached-prefix term it omits (0.13\% of the reference job). The \emph{gauge}, which sets the size of the unit, fixes one RIU as the decode of one token on a reference configuration: Llama-3.1-8B at FP8 with BF16 KV cache on one GPU, at context 2,048 and $n_0=32$. Measured on the evaluation node, this is 0.242~J of GPU board energy (Section~\ref{sec:hw}); the derived value of 0.34~J also counts an assumed host share.

The derivation fixes the structure of the coefficient set and leaves only values empirical. This has three consequences for the common billing form $\alpha\Tin+\beta T_{\mathrm{out}}+\gamma T_{\mathrm{rsn}}+\delta W_{\mathrm{multi}}$ over input, output, reasoning and multimodal quantities. Reasoning tokens are decoded like output tokens, so $\gamma=\beta$, as vendors already bill \cite{google2026gsu}; their counts stay on the record for audit because tenants cannot see them \cite{hoque2026inflation}. The multimodal term reduces to an encoder constant plus input tokens. The common form omits the KV cross term $\Tctx\Tgen$, which accounts for 42.0\% of the cost of a reference job (1,800 input tokens, 30\% cached, 420 output tokens on Llama-3.1-8B). A cached input token still carries $\mu\Tgen/(\alpha_1+\mu\Tgen)$ of a fresh token's cost, 0.64 at 420 output tokens; cache rates of 0.05 to 0.1 \cite{google2026gsu} shift this cost to the output price (derived; Supplement S1).

\subsection{Tokenisation: Native Counts and a Disclosed Premium Bound}
Let $\Phi_m(\ell)$ be the premium of language $\ell$ under model $m$'s tokeniser: the tokens needed for content in $\ell$ divided by the tokens needed for the same content in the reference language, on a public parallel corpus.
\begin{proposition}[Tokeniser-invariant billing]\label{prop:tok}
Suppose a meter charges by content alone and each model's rating is calibrated on the reference language, so that cost per native token does not depend on the language. If the models serving language $\ell$ span $\kappa_\ell=\max_m\Phi_m(\ell)/\min_m\Phi_m(\ell)$, the meter's relative error on $\ell$ is at least $(\kappa_\ell-1)/(\kappa_\ell+1)$ for some model, and an operator paid $p$ per reference token earns a margin $p-\lambda\Phi_m(\ell)$, negative whenever $\Phi_m(\ell)>p/\lambda$.
\end{proposition}
The best single charge is the harmonic mean of the extreme resources (Supplement S2). With the premium ranges of Fig.~\ref{fig:rq1}(a), the error bound is 14\% for Malay and 68\% for Tamil. Invariant billing is thus least accurate for the languages it is meant to protect, and without cross-subsidy operators have a reason to decline them. The architecture therefore meters native counts and publishes the premium as a bounded quality-class attribute, so a tenant declaring a language sees the eligible models before contracting. The framework therefore does not promise identical charges for equivalent content in different languages; it bounds and discloses the difference.

\subsection{Operator-Neutral Quantity and Operator Rating}
A settlement unit must give the same number on both sides of an operator boundary. Operators serving the same request may, however, spend different resources: published estimates for one model and query differ by a factor of 8.2 between two hosts \cite{jegham2025hungry}. If the operator is added to the ambient state, the argument of Theorem~\ref{thm:imposs} shows that a single rate cannot be both operator-neutral and cost-reflective. Coefficients are therefore keyed at two levels,
\begin{equation}\label{eq:twolevel}
U_v(q)=\langle\wv_{m,v},\phiv(q)\rangle,\qquad
\hat R_{o}(q)=\lambda_{o,c}\langle\wv_{m,v}^{(b)},\phiv(q)\rangle .
\end{equation}
The RIU, $U_v$, is the operator-neutral quantity that is aggregated, budgeted and settled, and it is the same for every operator and occupancy. The operator rating $\hat R_o$ estimates the resource behind it for cost recovery and, of the two, is the only one that depends on occupancy and configuration. It rescales the weight-stream coefficient, the only occupancy-dependent term of \eqref{eq:decode}, to the recorded occupancy band $b$, $\beta^{(b)}=\beta n_0/b$, and applies the efficiency $\lambda_{o,c}$ of configuration $c$. For MoE models, whose weight bytes per decode step grow with occupancy, the factor $n_0/b$ becomes $n_0W(b)/(b\,W(n_0))$, read from a measured weight-traffic table $W(b)$ (Section~\ref{sec:hw}); for dense models the two factors are identical. Where hardware differs in its ratio of compute to bandwidth, a configuration-specific coefficient vector replaces the scalar. The split also separates estimation: from latencies measured at controlled occupancy, a third party can recover coefficient ratios within the prefill group and within the decode group, but neither the level nor the ratio between groups (Supplement S2). The quantity's structure is thus externally checkable, while the operator rating must be declared and certified.

\subsection{Accuracy Classes and the Repackaging Bound}
Since Theorem~\ref{thm:imposs} rules out exact cost-reflectiveness, a meter declares its departure from it. For a workload $\mathcal{Q}$ the relative error is $e(\mathcal{Q})=(\sum_qU(q)-\lambda^{-1}R)/\lambda^{-1}R$.
\begin{definition}[Accuracy class]
A meter meets class $(\epsilon_b,\epsilon)$ at aggregation level $L$ on $\mathcal{E}$ if, over the reference workload, $|\mathbb{E}[e]|\le\epsilon_b$ and $\Pr(|e|\le\epsilon)\ge0.95$, with $e$ computed per request or per tenant statement (bill).
\end{definition}
Classes RIU-2, RIU-5, RIU-10 and RIU-25 are named by $\epsilon$ in per cent, with bias limits of 0.5, 1, 2 and 5\%. Per-request accuracy also bounds what a tenant gains by repackaging a request $q$ into another form $g(q)$:
\begin{theorem}[Repackaging bound]\label{thm:game}
If $|U(q)-\lambda^{-1}r(q)|\le\epsilon\lambda^{-1}r(q)$ holds for $q$ and for $g(q)$, any transformation $g$ with $r(g(q))\ge r(q)$ satisfies $U(g(q))\ge\frac{1-\epsilon}{1+\epsilon}U(q)$, so repackaging saves at most $2\epsilon/(1+\epsilon)$.
\end{theorem}
The proof combines the lower bound for $g(q)$ with the upper bound for $q$ (two-region form in Supplement S2). A class bounds only the 95th percentile, so a tenant who can aim at the tail is limited by the worst-case error. Disputes are therefore decided per statement: a correction is admissible only if it changes a statement by more than the declared $\epsilon$. Coefficient error is common to all requests of a statement and does not average out (Supplement S2), so bias is limited separately. A class is accepted on a one-sided test against the upper confidence bound, as for electricity meters \cite{iec62053-21}.

\subsection{Vintages and Re-Rating}
A \emph{vintage} fixes the reference workload, coefficients, envelope, accuracy class and effective and retirement dates. Values may change between vintages; the feature map and gauge may not, since changing them creates a new unit. Scaling every coefficient by one factor changes the gauge but not relative charges, so a uniform efficiency gain needs no recalibration; a change in coefficient ratios does.
\begin{proposition}[Re-rating]\label{prop:rerate}
For vintages $v,v'$ sharing the feature map and any set $\mathcal{Q}$ of requests on model $m$, $\sum_{q\in\mathcal{Q}}U_{v'}(q)=\langle\wv_{m,v'},\sum_{q\in\mathcal{Q}}\phiv(q)\rangle$. A scalar $T$ with $\sum U_{v'}=T\sum U_{v}$ for every $\mathcal{Q}$ in an envelope whose feature vectors span the feature space exists if and only if $\wv_{m,v'}=T\wv_{m,v}$.
\end{proposition}
The first statement is linearity of \eqref{eq:riu}; the second follows by evaluating single-request workloads (Supplement S2). A vintage that moves the reference occupancy changes $\beta$ relative to the other coefficients, so it admits no scalar conversion. Records carrying $\phiv(q)$, or its sum over a settlement bucket, re-rate exactly to any vintage that covers the model. Records of uncovered models wait for an extended vintage.

\section{SLA-Aware Utility Provisioning}\label{sec:prov}

\subsection{The Evaluation Node}
The evaluation setting is a shared node of four NVIDIA RTX 5000 Ada accelerators at UTAR; three of them were available for the measurements of Section~\ref{sec:hw}. Each card has 32~GB of GDDR6 memory with error correction, 576~GB/s of bandwidth, a 250~W board power limit and a PCIe~4.0~$\times$16 interface without NVLink \cite{nvidia_rtx5000ada_datasheet}. Error correction reserves 6.25\% of GDDR memory \cite{nvidia_cuda_best_practices}, leaving about 30~GB visible; the calibrated model budgets 27.44~GB after engine reservations. PCIe carries 31.5~GB/s per direction, 18 times less than local memory bandwidth, so models run as single-card replicas. Multi-instance GPU partitioning is supported on data-centre GPUs from Ampere onwards and on Blackwell workstation cards, but not on Ada cards \cite{nvidia_mig_supported_gpus}. Under NVIDIA's Multi-Process Service, a fatal fault in one client is reported to every client on that GPU \cite{nvidia_mps_when_to_use}. On this node every isolation guarantee is therefore a scheduler property, and, with more requested models than cards, which models stay loaded (residency) becomes an operational decision.

\subsection{Capacity and Quality Products}
Without a hardware partition, a capacity lease combines a scheduler \emph{floor} that reserves a share of each step's token budget, a \emph{KV-block reservation} that keeps other tenants' long contexts from blocking admission, and a \emph{residency reservation} that keeps the leased model loaded. The ratio of floors sold to sustainable throughput is published as a \emph{declared contention ratio}, as broadband access publishes its overbooking.

Five traffic classes, adapted from the contractual service categories of Asynchronous Transfer Mode (ATM) networks, carry the quality product: interactive guaranteed (IG) with tight time-to-first-token (TTFT) and time-per-output-token (TPOT) bounds, interactive bursty (IB), batch with a deadline (BB), adaptive (AD) and best effort (BE). A quality class is a conjunction of floors, $m\in\mathcal{K}\iff\bm{\theta}(m)\ge\underline{\bm{\theta}}^{\mathcal{K}}\wedge\mathrm{TTFT}_{99}\le\bar\tau^{\mathcal{K}}\wedge\Phi_m(\ell)\le\bar F^{\mathcal{K}}$ for every declared language, so classes form a partial order and no scalar ranking of models is needed. The router selects
\begin{equation}\label{eq:route}
\begin{aligned}
\pi(r)&=\arg\min\nolimits_{k\in\mathcal{A}(r)}\mathrm{cost}(r,k),\\
\mathcal{A}(r)&=\mathrm{LEGAL}(r)\cap\mathrm{CLASS}(r)\cap\mathrm{CONTRACT}(r),
\end{aligned}
\end{equation}
where $\mathrm{LEGAL}(r)$ holds the replicas a request may lawfully reach under data-residency, transfer-basis and export-control rules, evaluated against a registry at request time. Legality is a filter rather than a penalty, because a cheap enough forbidden route could outbid a penalty. The IDR stores the legal basis, which also meets the per-transfer record duty of Malaysia's cross-border framework \cite{cms_my_cbpdt_2025}. Residency changes per five-minute epoch rather than per engine step, because a cold load of a non-resident model takes 63--102~s on the node (Section~\ref{sec:hw}). Models that must be loaded cold are therefore pinned, and a model held in host memory can be swapped by sleep/wake in about 1~s. The controller loads a model only when pending work exceeds the work the swap loses by a factor $\tau$. It charges the swap to the operator, applies the legality predicate and counts refused requests as demand.

\subsection{Coupling Between Provisioning and Metering Accuracy}
The operator chooses a routing policy $\pi$, an admission policy $a$ and a declared envelope $\mathcal{E}$ to maximise SLO attainment subject to legality, class membership, tenant budgets $\sum_qU(q)\le b_i$ and
\begin{equation}\label{eq:couple}
\mathrm{supp}\;n(\pi,a)\subseteq\mathcal{E},
\end{equation}
where $n(\pi,a)$ is the occupancy distribution the two policies induce and $\mathrm{supp}$ is its support. Constraint \eqref{eq:couple} makes metering accuracy an output of the scheduler. The products are coupled through these declared parameters rather than a joint optimiser, because the unit changes allocation mainly as a budget (Section~\ref{sec:eval}).

\section{Case Study and Evaluation}\label{sec:eval}

\subsection{Setting, Data and Evidence}
The evaluation answers five research questions. RQ1: Do raw token counts measure inference work consistently across languages, tokenisers and batch occupancies? RQ2: How accurately do RIUs and occupancy-indexed ratings represent resource use across models and loads? RQ3: Which capacity and latency guarantees can a scheduler provide without hardware partitioning, and at what cost? RQ4: How does metering accuracy trade off against routing flexibility and service quality? RQ5 (Section~\ref{sec:fed}): How accurately can operators on different vintages settle? Table~\ref{tab:evidence} lists the evidence: token counts, architecture constants and the serving quantities of Section~\ref{sec:hw} are measured; costs computed from them and published specifications are derived; multi-tenant outcomes (RQ2--RQ5) come from the calibrated model.

\begin{table}[t]
\caption{Evidence used in the evaluation.}\label{tab:evidence}
\centering\footnotesize
\begin{tabular}{@{}p{2.35cm}p{3.3cm}>{\raggedright\arraybackslash}p{1.3cm}@{}}
\toprule
Source & Content & Class\\
\midrule
FLORES-200 devtest \cite{goyal2022flores101} & 1,012 parallel sentences, 18 languages, 16 tokenisers: 291,456 counts & measured\\
MultiPL-E \cite{cassano2023multiple} & 146 problems common to 24 programming languages, 10 tokenisers & measured\\
Model \texttt{config} files & layers, KV heads, windows, parameters of 13 models & measured\\
BurstGPT v1 \cite{wang2025burstgpt} & 1,429,737 request--response token pairs & trace\\
Azure trace 2023 \cite{azure_llm_trace_2023} & 19,366 conversation, 8,819 code requests & trace\\
Node runs E1--E8 (S7) & three RTX 5000 Ada, vLLM 0.11, NVML board energy & measured\\
Calibrated model (S4) & serving, routing, quality, settlement & calibrated model\\
\bottomrule
\end{tabular}
\end{table}

The calibrated model runs one continuous-batching engine per GPU; each decode round lasts the bytes moved (weights once, every sequence's KV state) divided by achieved bandwidth, plus admitted prefill time. Tenants draw request shapes from the two traces, and arrivals replay Azure inter-arrival times rescaled to a target load; each 1,800~s run serves 5,000 to 14,600 requests. Ground truth is a \emph{reference allocation}: a request is charged its own prefill, its own KV reads and an equal share of each weight read it joined. This is exact accounting under the model's chosen rule, not a uniquely correct economic allocation. The engine omits chunked prefill and per-request overhead. It uses the bandwidth and compute fractions, MoE weight traffic $W(n)$ and cold-load time measured in Section~\ref{sec:hw}, and an operator rating without $c_0$ (coefficient vintage RIU-2026.M); node power cancels in every reported ratio (earlier assumed constants: Supplement S7). The first study (RQ1, RQ2) serves Llama-3.1-8B at FP8, gpt-oss-20b at MXFP4, and Gemma-3-27B and Qwen3-32B at INT4 to five tenant classes over five arrival seeds. The second (RQ3, RQ4) uses eight single-card replicas on two logical operators of four cards each (the evaluation node and a second node in another jurisdiction), with eleven models over three seeds. Brackets give bootstrap intervals for measured quantities and the range over seeds for model results (details in Supplement S4).

\begin{figure*}[t]
\centering
\includegraphics[width=\textwidth]{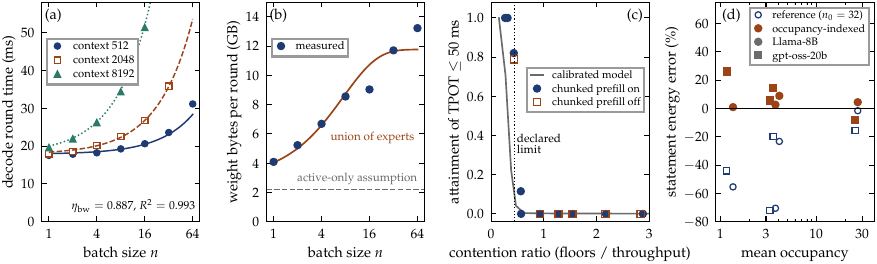}
\caption{Hardware validation on the evaluation node. (a) Decode round time against batch size, Llama-3.1-8B, with fitted roofline. (b) gpt-oss-20b weight bytes per decode round against batch size, with union-of-experts law and active-only read. (c) Guaranteed-class TPOT attainment against contention ratio, with the calibrated model. (d) Statement energy error against mean occupancy, reference and occupancy-indexed ratings. Measured.}\label{fig:hw}
\end{figure*}

\subsection{Hardware Validation}\label{sec:hw}
The node was measured first: three RTX 5000 Ada cards ran vLLM~0.11 with the four models of RQ2, with energy read from each card's NVML counter. Six decision rules were fixed before measurement (Supplement S7).

On the gauge model, a decode fit over 18 cells gives $\eta_{\mathrm{bw}}=0.887$ ($R^2=0.993$; Fig.~\ref{fig:hw}(a)), and prefill gives $\eta_{\mathrm{fl}}=0.41$. One gauge decode token costs 0.242~J of board energy at occupancy 32 and context 2,048. For gpt-oss-20b, weight bytes per decode round grow from 4.1~GB at $n=1$ to 13.2~GB at $n=64$ (panel (b)), because co-batched requests select different experts. The union-of-experts law $W(n)=2.94+8.83[1-(1-4/32)^n]$~GB fits within 19\% and replaces the constant active-weight read. The accelerator time per extra request is 0.65~ms on the gauge model, while a lone one-token request takes 18.5~ms of host and engine time, so $c_0$ moves from the operator rating to the gateway.

A cold load takes 63--102~s, so residency is pinned for cold models; vLLM sleep/wake restores a model held in host memory in 0.7--1.0~s. Under drawn leases with a 50~ms TPOT bound, guaranteed-class attainment is 1.00 at 0.31:1, 0.82 at 0.43:1 and 0.00 from 0.57:1 (panel (c)); the calibrated model, which omits chunked prefill, is more conservative and predicts 0.04 at 0.47:1. The paired cache contrast gives $d(420)=0.584$, against 0.641 derived.

On 199 MMLU-Pro items, accuracy separates the Q3 model (0.437) from the Q1 and Q2 models (0.628--0.719) but does not order Q1 and Q2, so class membership is declared from task-specific evaluation. An energy model fitted on controlled-occupancy cells and frozen was applied to eight 150~s trace runs (panel (d)). On the gauge model, the occupancy-indexed rating kept statement error within +9\% at mean occupancies of 1.4 to 27.1, while the reference rating erred by up to $-70.5$\%. For gpt-oss-20b, whose energy model keeps a constant weight stream, the error reached +26.2\% at occupancy 1.2, the MoE effect of panel (b). RQ2--RQ5 use a trace-driven model calibrated with these measurements, and RQ2 and RQ4 are reported at this measured vintage.

\subsection{RQ1: Raw Token Counts as a Measure of Work}

\begin{figure*}[t]
\centering
\includegraphics[width=\textwidth]{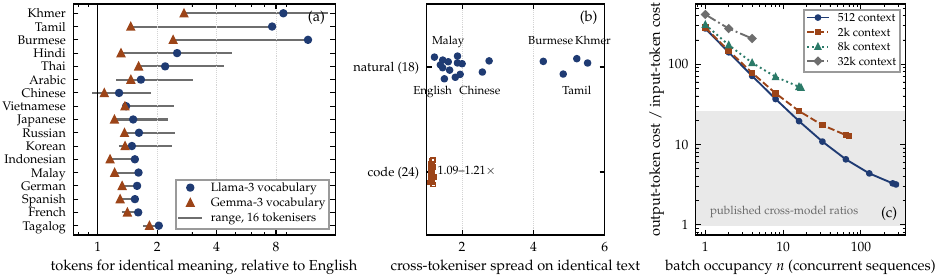}
\caption{Raw token counts as a measure of work. (a) Tokens for identical meaning relative to English, range over 16 tokenisers (measured, \mbox{FLORES-200} devtest; bootstrap intervals within markers). (b) Cross-tokeniser spread on identical text (measured). (c) Relative output-token cost by batch occupancy, Llama-3.1-8B at FP8 on one RTX 5000 Ada; curves end at KV capacity; band: vendor-published output-to-input ratios (derived).}\label{fig:rq1}
\end{figure*}

RQ1 establishes that raw token counts cannot serve as a common unit: counts vary with language and tokeniser for fixed content, and the work behind a fixed count varies with occupancy. Figure~\ref{fig:rq1}(a) shows that identical content yields different counts (Supplement S3). Malay needs 1.22 to 1.62 times the tokens of English; Tamil shows the widest range, 7.65 times [7.56, 7.73] under the Llama-3 vocabulary, against 1.47 times [1.45, 1.49] under Gemma-3. Panel (b) locates the effect: the spread across tokenisers on identical text is 1.22 times for English and 5.22 to 5.52 times for Tamil and Khmer, whereas every programming language lies between 1.09 and 1.21 times, so the disparity arises in natural-language vocabularies.

Panel (c) shows that equal counts can represent unequal work. On one model, card and context of 2,048 tokens, the relative cost of an output token falls from 284.8 at occupancy 1 to 12.65 at occupancy 70. This factor of 22.5 is fixed by published architecture and bandwidth alone, since the efficiency constants cancel. Across 648 combinations of memory fraction, error-correction reserve, precision and scheduler cap at contexts of 1,024 to 4,096 tokens the factor lies between 10.1 and 47.9; it shrinks at long contexts because KV capacity limits the batch (Supplement S3). One model's occupancy range thus exceeds the output-to-input ratios published across whole menus, 1 to 25 \cite{google2026gsu} and 4 to 8 \cite{azure2026ptusizing}.

\subsection{RQ2: Normalising Heterogeneous Consumption}

\begin{figure*}[t]
\centering
\includegraphics[width=\textwidth]{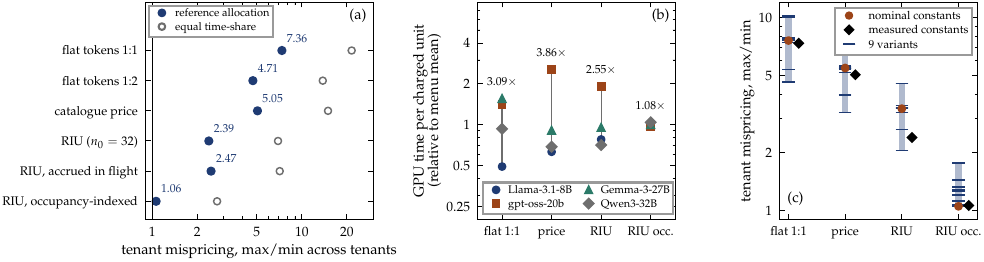}
\caption{Normalisation across models (calibrated model). (a) Tenant mispricing (bill share over GPU-time share, max/min across five tenants); bars: five seeds; filled: reference allocation; open: equal time shares. (b) GPU time per charged unit by model, relative to menu mean (load 1.0; markers: one seed; numbers: five-seed mean max/min). (c) Tenant mispricing under the earlier nominal constants, nine varied constants (meter derived at nominal values) and the measured constants.}\label{fig:rq2}
\end{figure*}

RQ2 establishes that recording occupancy, rather than calibrating a better price for each model, keeps the trued-up bill close to cost across models and loads. Six meters share one work-conserving weighted fair queue and differ only in what increments a tenant's counter: flat input and output counts weighted 1:1 and 1:2, the latter the default of VTC \cite{sheng2024fairness}; each model's catalogue input and output prices, as a multi-vendor aggregator does \cite{openrouter2026catalogue}; the RIU at reference occupancy, charged at admission or accrued during generation; and the RIU trued up with the occupancy-indexed operator rating of \eqref{eq:twolevel}. Against the reference allocation (Fig.~\ref{fig:rq2}(a)), flat 1:1 counting misprices by 7.36 times [7.01, 7.56], the catalogue price by 5.05 times, the RIU by 2.39 times [2.34, 2.46] and the trued-up bill by 1.06 times [1.03, 1.08], close to an oracle charging realised GPU time. The ordering also holds, with larger magnitudes, when each round's time is split equally (Supplement S4).

GPU time per charged unit across the four models (panel (b)) spreads 3.09 times under flat counting and 3.86 times under the catalogue price. The RIU at one reference occupancy narrows it to 2.55 times [2.39, 2.70]; the remainder arises because the replicas run at mean occupancies of 19.4 (Llama-3.1-8B), 11.5 (Qwen3-32B), 9.6 (Gemma-3-27B) and 5.3 (gpt-oss-20b) while the coefficients assume 32. Indexing by occupancy reduces the spread to 1.08 times; for the MoE model this uses the measured weight-traffic table, without which the spread is 1.79 times. Measured as excess above parity, a model-keyed table for one occupancy, as vendors publish, removes about a quarter of the spread of flat counting, whereas indexing by recorded occupancy removes 96\%; occupancy, rather than model identity, makes the charge cost-reflective. The strongest model-keyed alternative, a per-model rate table fitted to the node's own traffic, misprices by 1.07--1.19 times on four held-out seeds, close to the occupancy-indexed rating (1.03--1.07), but at loads of 0.6 or 1.5 it degrades to 1.48--1.66 times, while the occupancy-indexed rating stays at 1.07--1.08 (Supplement S5). Occupancy therefore belongs in the operator rating rather than in a one-off calibration.

The ordering held across nine variants of the assumed constants, with the measured constants inside their ranges (panel (c); Supplement S5).

\subsection{RQ3: Provisioning Without Hardware Partitioning}

RQ3 establishes that software isolation on shared accelerators can sell a throughput floor, while a latency target also requires a bounded contention ratio. In the calibrated model (Supplement S5, Fig.~S2(a)), routing freedom (in-class models per request) improved availability but not capacity. With each tenant restricted to its preferred model, 7.3\% of requests are refused because only eight of eleven models can be resident, and the worst tenant's TTFT attainment is 0.00. Two admissible models per request remove refusals but push displaced demand onto other tenants' replicas: throughput drops to 1,031~tok/s and the worst tenant's attainment stays at 0.003. All tenants recover with three or more models (0.89), reaching 1.00 at 1,680~tok/s under free in-class routing. Checking legality after the cost function would have sent 48.1\% of requests to a cheaper replica in a forbidden jurisdiction, while cross-operator routing changed throughput by only 0.5\%, so federation is a legal rather than a performance instrument. Among quality-of-service mechanisms, only the KV-block reservation changed an outcome, raising the bursty class from 0.20 to 0.63 (Supplement S5). Fig.~S2(b) shows that residency control pays off when swaps are fast: refusals fall from 7.3\% to zero at a 1~s swap, the measured sleep/wake time, and to 5.4\% at 10~s, whereas at the measured cold-load times they are 7.0--16.3\%.

Fig.~S2(c) quantifies the capacity product; the contention ratio is floors sold over aggregate throughput in the same run. Leases sold but not drawn leave the guaranteed class unchanged up to 5.5:1. When every lease draws its floor, the model predicts TPOT attainment of 1.00 at 0.15:1, 0.71 at 0.27:1 and 0.04 at 0.47:1, because a larger batch raises throughput but lengthens every decode round. Because the model omits chunked prefill, the prediction is conservative: on the node, attainment is 1.00 up to 0.31:1, 0.82 at 0.43:1 and 0.12 at 0.56:1 (Fig.~\ref{fig:hw}(c)). A lease with a latency target therefore needs a declared ratio below about 0.45:1, set by measurement, with the model as a safe planning bound. Software isolation thus provides a throughput floor and admission priority; the latency target also needs a low contention ratio.

\subsection{RQ4: Accuracy, Flexibility and Service}

\begin{figure*}[t]
\centering
\includegraphics[width=\textwidth]{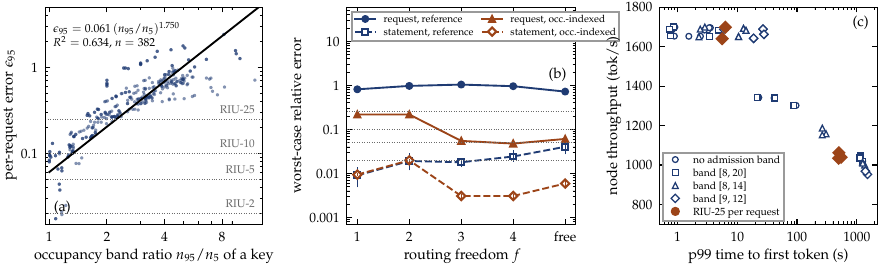}
\caption{Accuracy against flexibility and service (calibrated model). (a) Reference-meter per-request error against occupancy band ratio (382 (model, operator) keys) and fitted law. (b) Worst per-request and statement error of reference and occupancy-indexed meters by routing freedom (no admission control); bars: three seeds. (c) Throughput against p99 TTFT, 60 runs; filled: reference meter met per-request RIU-25 threshold.}\label{fig:rq4}
\end{figure*}

RQ4 establishes that recording occupancy delivers the billing accuracy that controlling occupancy buys only with throughput, as constraint \eqref{eq:couple} predicts and the separated design of Table~\ref{tab:designs} assumes. In Fig.~\ref{fig:rq4}(a), the reference meter is re-anchored to each (model, operator) key's median occupancy, with its rating fitted per key. Across 382 keys, its per-request error is $\epsilon_{95}=0.061\,(n_{95}/n_5)^{1.750}$ ($R^2=0.634$), where the band ratio measures the spread of occupancy. The fitted intercept is 6.1\% at a band ratio of one, against 13.4\% while the per-sequence constant $c_0$ remained in the operator rating. The tightest per-request class met in these runs was RIU-25, because the worst-served key sets the class: free routing strands some replicas near occupancy one, where the weight-stream term is largest. Meeting the per-request RIU-25 threshold required the tight admission band [9, 12]: at one model per request it cost 38\% of throughput, with queues that grow throughout the run (p99 TTFT of 496 to 547~s; panel (c)).

This per-request view overstates the error that matters for billing (panel (b)). Disputes concern statements, within which per-request noise averages while systematic error remains. For the reference meter, the worst tenant statement is in error by 0.5\% to 4.7\% across runs without admission control and the median by at most 2.0\%. Admission bands, by contrast, raise statement error to as much as 8.3\% by distorting which tenants share a batch. The occupancy-indexed rating, reading the occupancy recorded by the witness meter, kept the worst per-request error between 4\% and 23\% and the worst statement error at or below 2.2\% in all 60 runs without restricting routing or admission. With ratings fitted on one seed and frozen, the occupancy-indexed rating's worst statement error on 40 held-out runs was 0.1--2.1\% (reference meter 1.6--9.3\%; transfer across settings in Supplement S5). 

\subsection{Where Simpler Mechanisms Suffice}
Supplement S9 lists eleven configurations in which a simpler alternative performed as well as or better than a proposed mechanism, with the resulting design changes. They share a cause: without hardware partitioning, each guarantee that needed a narrow occupancy distribution was paid for in throughput. Under work-conserving scheduling every unit changed allocation only as a budget (Jain index within 0.0052); RIUs accrued in flight priced no better than RIUs charged at admission; quality-of-service mechanisms did not raise guaranteed-class TPOT attainment, because priority cannot reach a TPOT miss once a batch is formed; admission for per-request accuracy cost 38\% of throughput, whereas recording occupancy cost nothing; and residency control gave no benefit at the measured cold-load times, so cold models are pinned and swaps use sleep/wake.

\section{Regional Federation, Discussion and Limitations}\label{sec:fed}

\subsection{Federated Settlement (RQ5)}
Unlike roaming traffic, inference crosses an operator boundary mainly because of law or local saturation, since a token costs almost nothing to transport. Settlement borrows the machinery of roaming: an operator registry, a standard usage record, near-real-time exposure control and neutral clearing. The visited operator, which serves the request, validates entitlement with the home operator, which holds the tenant's balance; records are pushed home every $\Delta$ seconds, and a clearing service validates, re-rates and nets them multilaterally (Supplement S6). Between pushes, exposure is bounded by the visited operator's serving rate $C$, so $\Delta\le L_{\mathrm{r}}/C$ for a risk limit $L_{\mathrm{r}}$. Peak decode of the gauge model at its KV-limited occupancy of 70 on four cards gives $C\approx5.1\times10^3$~RIU/s, and $L_{\mathrm{r}}=10^5$~RIU then gives $\Delta\approx19$~s, against four hours for roaming call records \cite{gsma_nrtrde_2007}.

\begin{figure}[t]
\centering
\includegraphics[width=\columnwidth]{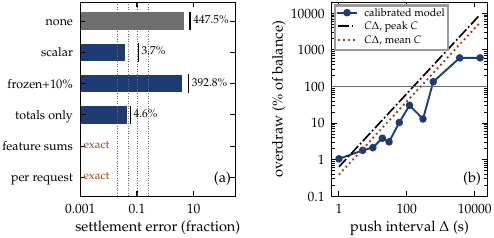}
\caption{Settlement across three logical operators on different coefficient vintages (calibrated model). (a) Settled-amount error against the same requests rated at the settlement vintage, 95th percentile (bar, with value) and maximum (tick mark) for: no re-rating, scalar factor, frozen coefficients with a 10\% haircut, bucket count and token totals, bucket feature sums, per-request feature vectors; dotted lines mark RIU-2 to RIU-25. (b) Maximum overdraw of a tenant opening 7,144 sessions across three operators against the push interval, with the bound $C\Delta$.}\label{fig:rq5}
\end{figure}

Three logical operators were modelled: one on a current vintage, one with a stale reference occupancy and one fitted before mixture-of-experts models joined the menu. Over 14,390 to 14,577 records per seed, net positions summed to at most $5.6\times10^{-17}$, showing only that records, re-rating and netting compose without loss. In Fig.~\ref{fig:rq5}(a) (two-seed means), settlement without re-rating errs by 448\% at the 95th percentile, and a frozen-coefficient fallback with a 10\% discount (haircut) by 393\% on a model whose cost structure changed, as Proposition~\ref{prop:rerate} predicts; a published scalar factor per model and vintage errs by 3.7\% but covers only the few records whose model kept its cost structure. Re-rating from per-request feature vectors or their per-bucket sums is exact by the same proposition, so exact settlement depends on records keeping the feature vector. Buckets keeping only request counts and token totals err by up to 5.6\% because the meter is non-linear in totals. Records of a model absent from the settlement vintage, nearly all cross-operator records in this deliberately adverse scenario, must be held until the vintage is extended. The home-side hold bounds rather than prevents overdraw by a tenant fanning out across operators (panel (b)): overdraw grows in steps with $\Delta$, from 1.1\% of the balance at 1~s and 4.4\% at 19~s to 603\% at four hours. The bound $C\Delta$ with the peak one-second rate holds from $\Delta=5$~s (Supplement S6).

\subsection{Deployment and Limitations}
Federation needs a neutral registrar for operator identifiers, vintages, sealed evaluation sets and clearing; ratings published under an agreed measurement boundary; and a scalable legal basis for cross-border routing (Supplement S6). A single university facility can deploy the single-operator subset without them. The multi-tenant studies (RQ2--RQ5) run on a trace-driven model, and the meter is derived from the same roofline model. The inputs that could change a decision are measured on the node (Section~\ref{sec:hw}), and a frozen calibration was checked on real traffic; that check covers eight runs of 150~s on two models, and its energy model for gpt-oss-20b still keeps a constant weight stream, which leaves errors of up to 26\%. The model omits chunked prefill, which makes its latency predictions conservative (Section~\ref{sec:eval}), and the RQ4 ratings were fitted in-sample, with held-out checks in Supplement S5. Energy is GPU board energy read from NVML, so measured joules exclude host and facility shares. The measurements cover one node with three cards, one accelerator generation and one engine version (vLLM~0.11); a new engine version or accelerator calls for re-verification under the rules of Supplement S7. The capacity charge is specified here and remains to be evaluated, settlement is verified in the calibrated model rather than between deployed operators, and a public benchmark separated the lowest quality class from the others but did not order the upper two, so class membership is declared from task-specific evaluation. The tokeniser study uses news-domain text and excludes two frontier vocabularies that are not public. Both meters sit inside one operator's trust boundary, so auditability depends on an external auditor and on substitution and count audits that exist as detectors but are not yet deployed as services \cite{zhang2026iris}.

\section{Conclusion}\label{sec:concl}
Raw token counts are not a model-independent measure of inference work: identical content needs 1.2 to 1.6 times more tokens in Malay than in English and up to 7.65 times more in Tamil, and at common context lengths output-token cost varies 10 to 48 times with batch occupancy. Under batch sharing, a per-request meter cannot be additive, ambient-independent and exactly cost-reflective at once. Token-as-a-Utility therefore assigns these properties to separate products and recovers residual cost through an occupancy-indexed operator rating. Measurements on the evaluation node confirmed the efficiency, cache-cost and contention assumptions of the design, and on real traffic the occupancy-indexed rating kept the gauge model's statement energy error within 9\%. In the calibrated model, RIUs reduced tenant mispricing from 7.36 times under flat tokens to 2.39 times, and the occupancy true-up reduced it to 1.06 times and stayed accurate on held-out runs and under load shifts. Records carrying the feature vector re-rate exactly across operators and vintages. A per-token latency target held only below a measured contention ratio of about 0.45:1, and accuracy classes are declared per statement. The measurements also added a weight-traffic table for mixture-of-experts models, a gateway-level per-request constant and pinned residency for cold models. A deployed federation awaits the institutions identified above.

\section*{Data Availability and Acknowledgments}
Token counts, model constants, hardware measurements and scripts (E1--E8), and the calibrated-model code and outputs are archived on Zenodo, \url{https://doi.org/10.5281/zenodo.23792887}. The concepts, framework, study design and interpretation are the authors' own. Claude (Anthropic) was used as an assistive tool for code development, data processing and language editing under the authors' direction; the authors reviewed every analysis and passage, verified all results, and take full responsibility for the content.

\bibliographystyle{IEEEtran}
\bibliography{references}

\end{document}